\documentclass[aps,prb,twocolumn,showpacs,superscriptaddress]{revtex4}
\usepackage{graphicx}
\usepackage{bm}

\newcommand{\beq}{\begin{equation}}
\newcommand{\eeq}{\end{equation}}

\newcommand{\fl}{\mbox{\scriptsize FL}}

\begin{document}

\title{
Fermi surface reconstruction in strongly correlated Fermi systems
as a first order phase transition}
\author{S.~S.~Pankratov}
\affiliation{Russian Research Centre Kurchatov Institute, Moscow,
123182, Russia} \affiliation{Moscow Institute of Physics and
Technology, Moscow, 123098, Russia}
\author{M.~V.~Zverev}
\affiliation{Russian Research Centre Kurchatov Institute, Moscow,
123182, Russia} \affiliation{Moscow Institute of Physics and
Technology, Moscow, 123098, Russia}
\author{M.~Baldo}
\affiliation{Istituto Nazionale di Fisica Nucleare, Sezione di Catania, 64
Via S.-Sofia, I-95123 Catania, Italy}

\date{\today}

\begin{abstract} A quantum phase transition in strongly correlated
Fermi systems beyond the topological quantum critical point is
studied within the Fermi liquid approach. The transition occurs
between two topologically equivalent states, each with three sheets
of the Fermi surface. One of these states possesses a quasiparticle
halo in the quasiparticle momentum distribution $n(p)$, while the
other, the hole pocket. The transition is found to be of the first
order with respect to both the coupling constant $g$ and the
temperature $T$. The phase diagram of the system
in the vicinity of this transition is constructed.

\end{abstract}


\maketitle

Low temperature quantum phase transitions in strongly correlated
Fermi systems is one of hot topics in the condensed matter physics
in the last decade. Variation of external parameters (pressure,
density, magnetic field) allows one to shift the transition
temperature to zero and to obtain the quantum critical point, which
is associated with divergence of the effective mass $M^*$. In the
vicinity of this point, low temperature properties of the system
possess non-Fermi-liquid character, i.e.\ they are not described
within the conventional Landau theory of Fermi liquid.

At present, experimental information on the quantum critical point
is available only for three types of strongly correlated Fermi
systems: i) the inversion layer in MOSFET silicon transistors in
which electrons form a two-dimensional (2D) liquid,
\cite{Shashkin-PRB-2002,Pudalov-PRL-2002} ii) films of $^3$He atoms
on various substrates, \cite{Godfrin-JLT-1998,Saunders-Science-2007}
iii) metals with heavy fermions.
\cite{Oeschler-PB-2008,Gegenwart-NP-2008}

In nonsuperfluid homogeneous and isotropic Fermi systems, which will
be considered in this work, the ratio of the effective mass $M^*$ to
the bare one $M$ reads \beq {M\over M^*}=
z\left[1+\left({\partial\Sigma(p,\varepsilon)\over\partial\epsilon^0_p}
\right)_0\right] \ , \label{meffz} \eeq where
$\epsilon^0_p=p^2/2M-\mu$, $\mu$ is the chemical potential, $\Sigma$
is the mass operator, and the quasiparticle weight $z$ in a single
particle state is given by
$z=[1-\left(\partial\Sigma(p,\varepsilon)/
\partial\varepsilon\right)_0]^{-1}$ (index $0$ means evaluation of the derivative on the Fermi surface).
The formula (\ref{meffz}) allows one to consider two scenarios of
the quantum critical point. The collective scenario is build on a
supposition that energy dependence of the mass operator prevails
over its momentum dependence due to exchange by critical
fluctuations in the vicinity of collapse point of the respective
collective mode, and leads to vanishing of the quasiparticle weight
$z$ and, hence, to divergence of the effective mass $M^*$ just at
that point.
\cite{Gegenwart-NP-2008,Hertz-PRB-1976,Millis-PRB-1993,Coleman-JPCM-2001}
The topological scenario of the critical point assumes $z$-factor to
be finite at that point, however dominating momentum dependence of
the mass operator results in the change of the Fermi surface
topology.
\cite{KS-JETPL-1990,Volovik-JETPL-1991,Nozieres-1992,Khodel-JETPL-2007,KCZ-PRB-2008}
The reader can find comparison of these two scenarious in
Refs.~\onlinecite{Khodel-JETPL-2007,KCZ-PRB-2008}. In this paper, we
consider the topological scenario of the quantum critical point.

In this connection, it is worth to note that in accordance with
topological classification \cite{Volovik-book-2007} of ground states
of fermionic systems, the basic classes differ by topological
dimension ${\cal D}$ of the manifold of nodes of the single-particle
spectrum $\epsilon(p)$ measured from the chemical potential. Within
the same class, we will distinguish states by a number of connected
sheets of that manifold. All transitions between ground states which
belong to different topological classes or transitions between
states with different topology
in the same class are quantum phase transitions occurring at $T=0$.
Conventional nonsuperfluid homogeneous and isotropic Fermi liquid at
$T=0$ with quasiparticle momentum distribution
$n_{\fl}(p)=\theta(p_F-p)$ belongs to the class for which the
dimension ${\cal D}$ of the manifold of nodes is less by unity than
the dimension of the system itself, and this manifold is a single
connected sheet, i.e. the Fermi surface.

Violation of the necessary stability condition for Landau quasiparticle ground state
with the momentum distribution $n_{\fl}(p)$ serves a signal for its
topological reconstruction. This stability condition \beq \delta E
=2\int \epsilon (p,[n_{\fl}(p)])\, \delta n_{\fl}({\bf p}) \,
d\upsilon>0 \label{necessary} \eeq demands positivity of variation
of the ground state energy $E$  for any admissible variation $\delta
n_{\fl}({\bf p})$ which satisfies the condition \beq 2\int\delta
n_{\fl}({\bf p})\,d\upsilon = 0\ . \label{norm} \eeq In
Eqs.~(\ref{necessary}) and (\ref{norm}), $d\upsilon$ denotes an
elementary volume of the momentum space, and a factor of two
means summation over two spin projections. The distribution $n_{\fl}(p)$
satisfies the necessary condition (\ref{necessary}) provided the
single-particle spectrum $\epsilon(p,[n_{\fl}(p)])$ vanishes only at
$p=p_F$. In weakly and moderately correlated systems this is true.
However, in process of correlations strengthening with the change of
external parameters, new nodes
of the function $\epsilon(p,[n_{\fl}(p)])$ can appear and the
condition (\ref{necessary}) is then violated.
\cite{KS-JETPL-1990,Volovik-JETPL-1991,Nozieres-1992}

\begin{figure}[t]
\hskip 0.5 cm
\includegraphics[width=0.8\linewidth,height=0.7\linewidth]{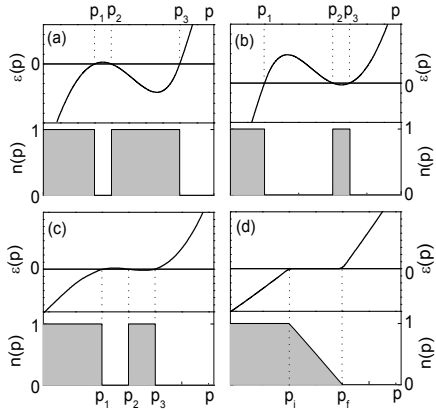}
\caption{Single-particle spectrum and momentum distribution of
quasiparticles: illustration of topological reconstruction
scenarios. Panel (a): hole pocket, panel (b): quasiparticle halo,
panel (c): symmetrical three-connected distribution, panel (d):
fermion condensate. } \label{fig:scenarios}
\end{figure}

Two scenarios of topological reconstruction of the momentum
distribution $n_{\fl}(p)$ resulting from this violation are known:
i) reconstruction within the same topological class, i.e.\ with no
change of the dimension ${\cal D}$ but with change of the number of
connected sheets of the Fermi surface,
\cite{Frohlich-PR-1950,Vary-PRC-1979,Zabolitsky-PRC-1979,Aguilera-PRC-1982,
ZB-JETP-1998,ZB-JPCM-1999,Shaginyan-JETPL-1998,Schofield-PRB-2006}
ii) reconstruction with transition to another topological class,
i.e.\ with change of the dimension ${\cal D}$.
\cite{KS-JETPL-1990,Volovik-JETPL-1991,Nozieres-1992} When new nodes
of the spectrum appear on the same side from the Fermi surface, the
distribution $n_{\fl}(p)$ is rearranged to asymmetric
three-connected momentum distribution which is schematically shown
in the panels (a) and (b) of Fig.~\ref{fig:scenarios}. If the nodes
$p_1$, $p_2$, $p_3$ are arranged in such a way that $|p_1-p_2| \ll
|p_2-p_3|$ the distribution has a form of the {\it hole pocket} in
the filled sphere (we refer to this state as to ${\cal H}$-state),
while if $|p_1-p_2| \gg |p_2-p_3|$, one deals with the {\it
quasiparticle halo} (${\cal P}$-state). Such reconstruction results
in no change of the topological dimension ${\cal D}$,
while the number of connected sheets of the Fermi surface appears to equal three.
If new nodes of the spectrum emerge to both sides of the Fermi
surface, then together with formation of symmetrical three-connected
Fermi surface
\cite{ZB-JETP-1998,ZB-JPCM-1999,Shaginyan-JETPL-1998}
(see panel (c) of Fig.~\ref{fig:scenarios}), essentially different
scenario of rearrangement of Landau state is possible,
fermion condensation,
\cite{KS-JETPL-1990,Volovik-JETPL-1991,Nozieres-1992}
shown in the panel (d) of Fig.~\ref{fig:scenarios}. In this scenario the
quasiparticle momentum distribution gradually drops within the
interval $p_i<p<p_f$, and the spectrum $\epsilon(p)$ identically
vanishes within this interval. Hence the state with fermion
condensate turns out to belong to the class with the topological
dimension of manifold of nodes ${\cal D}$ coinciding with dimension
of the system. The fermion condensate, revealed
and studied in details about 20 years ago,
\cite{KS-JETPL-1990,Volovik-JETPL-1991,Nozieres-1992} acquires a new
life in these days in a form of topologically protected flat bands,
i.e.\ dispersionless branches of the single-particle spectrum with
exactly zero energy. \cite{Volovik-1012-0905,Brydon-1104-2257}
Particularly, possibility of existence of surface states with flat
band is intensively discussed,
\cite{Volovik-1012-0905,Brydon-1104-2257,Schnyder-1011-1438,Volovik-1103-2033}
which may be superconducting with high transition temperature.
\cite{Volovik-1103-2033}

In this paper, we consider the scenario of the topological
reconstruction with formation of three-connected Fermi surface. We
will show that in a topologically rearranged system,
the first order transition between ${\cal P}$- and ${\cal H}$-states may
occur.

We focus now on the scenario of topological transition in which only
regions adjacent to the Fermi surface are involved. Results of
microscopic calculations for 2D liquid $^3$He
\cite{Krotscheck-PRL-2003} and for low-density 2D electron gas
\cite{BZ-JETPL-2005}
indicate this way of topological reconstruction in these systems. For
evaluation of the single-particle spectrum $\epsilon(p)$ and
momentum distribution of quasiparticles $n(p)$ we use the
Fermi-liquid relation
\cite{Landau-JETP-1956,Landau-JETP-1958,LL-Stat-IX,Trio} \beq
{\partial\epsilon(p)\over\partial {\bf p}} =
  {{\bf p}\over M} +
     \int\! f({\bf p},{\bf p_1})\,
         {\partial n(p_1)\over\partial {\bf p_1}}\, d\upsilon_1 \ ,
\label{lansp}
\eeq in which $n(p) = \theta(-\epsilon(p))$ and the
quasiparticle interaction $f({\bf p},{\bf p_1})$ in the Fermi liquid
theory is supposed to be known function of momenta. The formula
(\ref{lansp}) represents
the nonlinear integro-differential equation for the single-particle
spectrum $\epsilon(p)$. Any numerical algorithm of its solution
requires use of regularization procedure. This is finite temperature
that plays a role of a natural physical regularizer. Indeed, making
use of the Fermi-Dirac relation between momentum distribution and
spectrum \beq n(p,T)=\left[1+e^{\epsilon(p)/T}\right]^{-1}
\label{dist} \eeq allows one to solve Eq.~(\ref{lansp}) by standard
iterative algorithm.

\begin{figure}[t]
\centerline{\includegraphics[width=0.95\linewidth,height=0.85\linewidth]{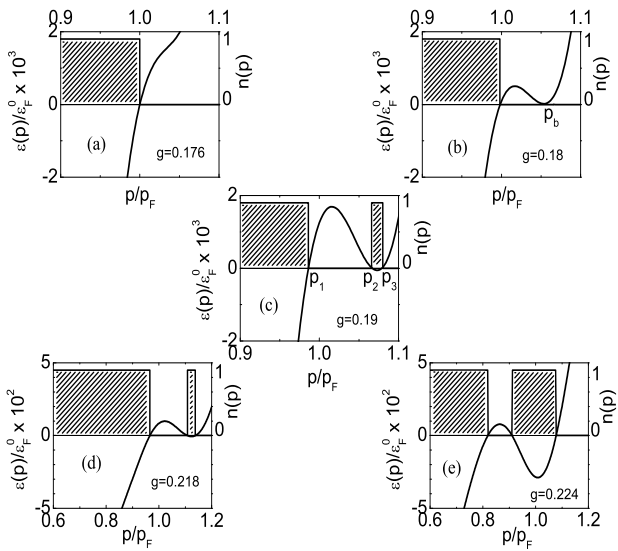}}
\caption{Single-particle spectra $\epsilon(p)$ and quasiparticle
momentum distributions $n(p)$ evaluated at $T=10^{-5}$ in units of
$\varepsilon_F^0=p_F^2/2M$ for the model with the interaction
(\ref{model_2pf}) with $q_0=2p_F$ for different values of the
interaction constant $g$: 0.176 (panel (a)), 0.180 (panel (b)),
0.190 (panel (c)), 0.218 (panel (d)) and 0.224 (panel (e)). }
\label{fig:spectra}
\end{figure}

We analyze the topological reconstruction in 2D Fermi system with a quasiparticle interaction function
\beq
f({\bf p},{\bf p_1})=-g{\pi\over
M} {1\over (({\bf p}-{\bf p_1})^2/q_0^2-1)^2+\beta^2} \ ,
\label{model_2pf}
\eeq
with $q_0\simeq 2p_F$, $\beta$=0.14, which enables one to reproduce adequately microscopic calculations
     \cite{BZ-JETPL-2005}
of single-particle spectra of 2D electronic gas at $T=0$ on the
Fermi-liquid side from the quantum critical point.
Since the interaction function depends on the difference ${\bf p}-{\bf p}_1$,
Eq.~(\ref{lansp}) is integrated to the form \beq \epsilon(p) =
{p^2\over 2M} -\mu +
     \int\! f({\bf p},{\bf p_1})\,
         n(p_1)\, d\upsilon_1 \ ,
\label{lansp1} \eeq in which the chemical potential $\mu$ is
obtained from the normalization condition \beq 2\int n(p)\,
d\upsilon = \rho \ . \label{norm1} \eeq Single-particle spectrum
$\epsilon(p)$ and quasiparticle momentum distribution are evaluated
by self-consistent solution of Eqs.~(\ref{dist}), (\ref{lansp1}) and
(\ref{norm1}). Rearrangement of the ground state of the considered
system with increase of the interaction constant $g$ is shown in
Fig.~\ref{fig:spectra}. Calculations are performed at
  $T=10^{-5}\varepsilon_F^0$
modeling zero temperature. Irregularity of
the spectrum at $p>p_F$ distinguished in the panel (a) of this
Figure is developed to its nonmonotonous behavior which, as $g$
reaches $g_b=0.18$, results in the bifurcation in the equation $\epsilon(p)=0$ at
$p=p_b>p_F$ (see panel (b)) and then, to the topological
reconstruction with formation of the ${\cal P}$-state (panel (c)).

\begin{figure}[t!]
\centerline{\includegraphics[width=0.9\linewidth,height=0.45\linewidth]{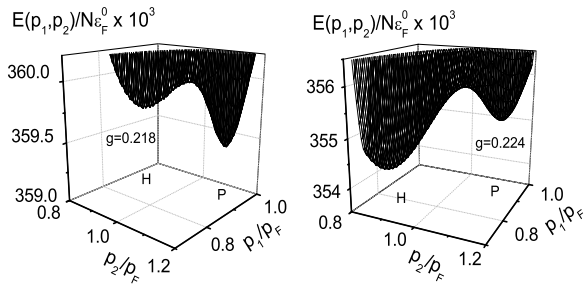}}
\caption{Energy per one particle as a function of $p_1$ and $p_2$ evaluated at $T=0$ for $q_0=2\,p_F$.
}
\label{fig:minima}
\end{figure}

The three-connected momentum distribution at zero temperature
$n_3(p)=\theta(p_1-p)-\theta(p_2-p)+\theta(p_3-p)$ is determined in
the functional space by two independent parameters, the third one
being obtained from the relation \beq p_1^2-p_2^2+p_3^2=p_F^2 \ ,
\label{norm_123} \eeq following from the normalization condition
(\ref{norm1}). This implies that the energy functional of the system
\beq E[n] = 2\int\frac{p^2}{2M}n({\bf p})\,d\upsilon + \int f({\bf
p},{\bf p_1})n({\bf p})n({\bf p_1})\,d\upsilon d\upsilon_1
\label{energy} \eeq considered within the class of distributions
$n_3(p)$ is just a function of two variables, say, $p_1$ and $p_2$.
Evaluation of the function $E(p_1,p_2)$ indicates that the momentum
distribution $n(p)$ obtained by self-consistent solution of
Eqs.~(\ref{dist}), (\ref{lansp1}), (\ref{norm1}) and shown in the
panel (c) of Fig.~\ref{fig:spectra} corresponds to the global
minimum of this function. There is no other local minimum of
$E(p_1,p_2)$ just beyond the topological transition point. However,
the situation changes with increasing coupling constant, namely, a
new minimum appears at $g\simeq 0.21$. The relief of the function
$E(p_1,p_2)$ at $g=0.218$ is shown in the left panel of
Fig.~\ref{fig:minima}. The deep minimum at $p_2>p_F$ corresponds to
the ground ${\cal P}$-state, the quasiparticle momentum distribution
and the spectrum of which are shown in the panel (d) of
Fig.~\ref{fig:spectra}. The shallow minimum at $p_2<p_F$
corresponds to the
metastable ${\cal H}$-state which is obtained by solving of the set
of Eqs.~(\ref{dist}), (\ref{lansp1}), (\ref{norm1}) provided
the iteration procedure is started from a state inside the shallow well
 With further increasing of the coupling constant $g$, the ${\cal H}$-state
minimum lowers with respect to the ${\cal P}$-state minimum, and both
minima equalize at $g=0.22$. The first-order transition from
the ${\cal P}$-state to the ${\cal H}$-state occurs at this point,
the latter state becomes the ground one at $g>0.22$. This is
demonstrated in the right panel of Fig.~\ref{fig:minima} where the
relief $E(p_1,p_2)$ at $g=0.224$ is drawn. The deep minimum
corresponds to the ${\cal H}$-state shown in the panel (c) of
Fig.~\ref{fig:spectra}. As follows from calculations made up to
$g=0.26$, the ${\cal P}$-state keeps on existing as a metastable
one.

\begin{figure}[h]
\centerline{\includegraphics[width=0.63\linewidth,height=0.45\linewidth]{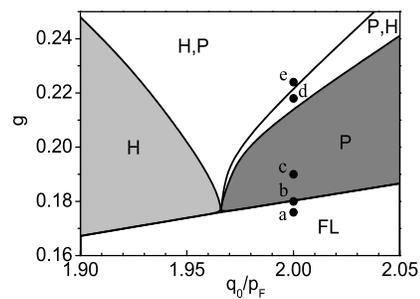}}
\caption{Phase diagram in $(q_0,g)$ variables.
Capital letters denote the state occupying corresponding part of the diagram. Two
letters mean that the first state is the ground state, while the
second one is metastable. } \label{fig:phase_dia}
\end{figure}

\begin{figure}[h]
\centerline{\includegraphics[width=0.9\linewidth,height=0.67\linewidth]{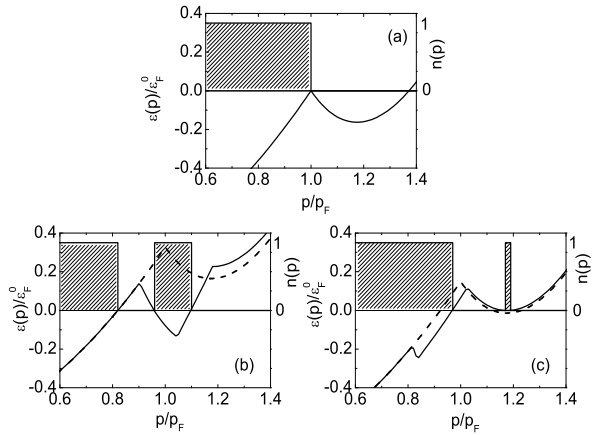}}
\caption{Quasiparticle momentum distributions and single-particle
spectra evaluated in the model (\ref{delta_model}). Panel (a) shows
unstable Landau state, panel (b) demonstrates the ${\cal H}$-state,
panel (c) --- the ${\cal P}$-state. Single-particle spectrum for the
Landau state $\epsilon(p,[n_{\fl}])$ shifted by the respective
difference of chemical potentials is shown for comparison in panels
(b) and (c) by dashed lines. } \label{fig:model_spec}
\end{figure}

\begin{figure}[h]
\centerline{\includegraphics[width=0.6\linewidth,height=0.5\linewidth]{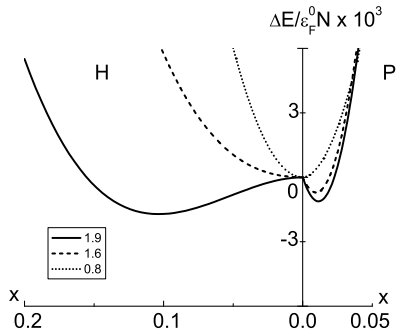}}
\caption{Energies per one particle in units of $\varepsilon^0_F$
evaluated for the ${\cal H}$-state (left part of the figure) and the
${\cal P}$-state (right part) at three values of the constant $f$. }
\label{fig:model_energy}
\end{figure}

Analysis of metamorphoses of solutions of Eqs.~(\ref{dist}),
(\ref{lansp1}), (\ref{norm1}) with variation of both the coupling constant
$g$ and the wave vector $q_0$ allows one to build the phase diagram
of the system in these variables which is shown in
Fig.~\ref{fig:phase_dia}. At $q_0>1.96\,p_F$, the diagram is
arranged similarly to the one considered above for $q_0=2\,p_F$.
Five points on it correspond to five solutions shown in
Fig.~\ref{fig:spectra}. Arrangement of the phase diagram at
$q_0<1.96\,p_F$ is different, namely, the three-connected ${\cal
H}$-state emerges just beyond the point of the topological
transition and remains the ground state while the metastable ${\cal
P}$-state appears with increasing $g$.

It is worth noting that the first-order transition under
consideration is not inherent in 2D systems only.
Analysis for 3D systems shows that an analogous transition occurs
in 3D as well.

Why the considered set of equations possesses simultaneously two
solutions at fixed parameters, can be understood with a help of a
simplified model with $\delta$-function quasiparticle interaction.
3D system is somewhat more convenient for this purpose than 2D one
since all calculations can be done analytically for the 3D case. For
the model interaction \beq f(q) = -f{8\pi^2p_F\over M}\delta({\bf
q}^2-q_0^2) \label{delta_model} \eeq with $q_0\simeq2\,p_F$, the
state with quasiparticle momentum distribution $n_3(p)$ at $T=0$ has
the spectrum
\begin{eqnarray}
\lefteqn{ \epsilon(p;[n_3]) = {p^2\over 2M} -\mu } \nonumber \\
 & &
 +f{p_F\over 2Mp}
\sum_{k=1}^3(-1)^{k+1}[(p{-}q_0)^2{-}p_k^2]\,\theta(p{+}p_k{-}q_0).
\qquad
\label{spectr_3}
\end{eqnarray}
Single-particle spectra evaluated with use of Eq.~(\ref{spectr_3})
with $q_0=2\,p_F$, $f=2.0$ are displayed in
Fig.~\ref{fig:model_spec}. The spectrum $\epsilon(p;[n_{\fl}])$
given by an account of the only term in the sum in (\ref{spectr_3})
with the boundary momentum $p_F$ is shown in the panel (a). Due to
$\theta$-function on the r.h.s.\ of Eq.~(\ref{spectr_3}), the
spectrum possesses a kink and
changes its behavior at the point $p^{(1)}=q_0-p_F=p_F$. The necessary
condition (\ref{necessary}) for stability of the Landau state with
the quasiparticle distribution $n_{\fl}(p)$ is, evidently, violated.
The spectra $\epsilon(p;[n_3])$ shown by solid lines on panels (b)
and (c) possesses three kinks. If $p_2<p_F$, the second kink is
placed at the point $p^{(2)}=q_0-p_2$ lying to
the right of $p_F$. In this case tuning of
the chemical potential to the condition of conservation of the
quasiparticle number gives rise self-consistently to the ${\cal
H}$-state. Such state with the nodes $p_1=0.82\,p_F$,
$p_2=0.96\,p_F$, $p_3=1.1\,p_F$ is shown on panel (b) of
Fig.~\ref{fig:model_spec} together with the spectrum for the Landau
state shifted for convenience by the difference of the chemical
potentials. In case $p_2>p_F$, the point of the second kink is
placed to the left of $p_F$, and then tuning of
the chemical potential gives rise to the ${\cal P}$-state. The
spectrum of this state with the nodes $p_1=0.97\,p_F$,
$p_2=1.16\,p_F$, $p_3=1.18\,p_F$ is shown on panel (c) together with
shifted spectrum $\epsilon(p;[n_{\fl}])$.

To elucidate which of the two states, ${\cal H}$ or ${\cal
P}$, proves to be the ground one, we evaluate the energies of these
states. Dimensionless energy of the tree-connected state per one
particle ${\cal E}=\Delta E/\varepsilon_F^0N$ measured from the
energy of the Landau state is given as follows
\begin{eqnarray}
\lefteqn{ {\cal E} = \frac{3}{5 p_F^5}
\left(p_1^5-p_2^5+p_3^5-p_F^5\right) } \qquad\qquad
\nonumber\\
&&-\frac{2fq_0}{p_F}\left(S(q_0;[n_3])-S(q_0;[n_{\fl}])\right). \qquad
\label{energy_3}
\end{eqnarray}
Upon not difficult but cumbersome algebra, the structure function
\beq S(q;[n]) = \frac{2}{\rho}\int n({\bf p}+{\bf q})n({\bf p})
\frac{d^3{\bf p}}{(2\pi)^3} \label{struct} \eeq is evaluated
analytically. Excluding, say, the variable $p_3$, one then arrives
at the energy as a function of two variables, $p_1$ and $p_2$. The
condition of its extremum allows one to express $p_2$ via $p_1$ and
reduce the energy to the function of a single variable $p_1$. Let
$q_0=2\,p_F$, we introduce then a new convenient variable
$x=1-p_1/p_F$. For small values of $x$, the energy of the ${\cal
P}$-state equals \beq {\cal E}_{\cal P}(x) = -a_{\cal
P}\,(\delta_{\cal P}^2-\frac{3}{4}\delta_{\cal
P}^3)\,\theta(\delta_{\cal P})\,x + b_{\cal P}\,x^2,
\label{energy_p} \eeq where $\delta_{\cal P} = f - f_{\cal P}^c$ is
an excess of the coupling constant $f$ over the critical value
$f_{\cal P}^c=1$ corresponding to the topological transition from
the Landau state to the ${\cal P}$-state, $a_{\cal P}=3/4$, $b_{\cal
P}=6$. For the ${\cal H}$-state, one analogously obtains \beq {\cal
E}_{\cal H}(x) = -a_{\cal H}\,\delta_{\cal H}\,x^2 + b_{\cal
H}\,x^3, \label{energy_h} \eeq where $\delta_{\cal H}=f-f_{\cal
H}^c$, $f_{\cal H}^c=1+1/\sqrt{2}$ is a critical value of the
constant at which the metastable ${\cal H}$-state emerges, $a_{\cal
H}=6(\sqrt{2}-1)$, $b_{\cal H}=9/2-\sqrt{2}$. Expressions
(\ref{energy_p}) and (\ref{energy_h}), formally applicable in the
vicinity of the respective critical constants, qualitatively
describe behavior of the energies of ${\cal P}$- and ${\cal
H}$-phases far from $f_{\cal P}^c$ and $f_{\cal H}^c$ as well.

As long as $\delta_{\cal P}<0$, the linear in $x$ term in the energy
excess of the ${\cal P}$-state over the Landau state equals zero and
${\cal E}_{\cal P}(x)=b_{\cal P}\,x^2>0$. For such values of the
coupling constant, $\delta_{\cal H}$ is also negative and, hence,
both terms in ${\cal E}_{\cal H}(x)$ are positive. Therefore,
the Landau state with the quasiparticle distribution $n_{\fl}(p)$ is
the ground state. The functions
${\cal E}_{\cal P}(x)$ and ${\cal E}_{\cal H}(x)$
at $f<f_{\cal P}^c$ are shown by dotted
lines in Fig.~\ref{fig:model_energy}. At $\delta_{\cal P}>0$, linear
in $x$ term with minus sign emerges in ${\cal E}_{\cal P}(x)$. As a
result, the ${\cal P}$-state wins the contest against the Landau
state. Thus, the second order topological transitions occurs at
$f=f_{\cal P}^c$, namely, the three-connected ${\cal P}$-state with
the quasiparticle halo appears to be the ground state. The energy
${\cal E}_{\cal H}(x)$ remains a monotonically growing function of
$x$ as long as $\delta_{\cal H}<0$. Both energy curves at the
constant $f=1.6$ corresponding to the case $f_{\cal P}^c<f<f_{\cal
H}^c$ are shown by dashed lines in Fig.~\ref{fig:model_energy}. As
soon as $f$ exceeds $f_{\cal H}^c$, the coefficient near the
quadratic term in the function ${\cal E}_{\cal H}(x)$ changes the
sign and the function acquires the minimum, i.e.\ the metastable
${\cal H}$-state appears. When with incresing $f$, this minimum
shown by a solid curve in the left part of
Fig.~\ref{fig:model_energy} becomes deeper than the minimum of the
right solid curve ${\cal E}_{\cal P}(x)$, the three-connected ${\cal
H}$-state with the hole pocket, topologically equivalent to the
${\cal P}$-state, becomes the ground state. The transition between
the ${\cal P}$- and ${\cal H}$-states is first order.

\begin{figure}[t]
\vskip 0.5 cm
\centerline{\includegraphics[width=0.9\linewidth,height=0.7\linewidth]{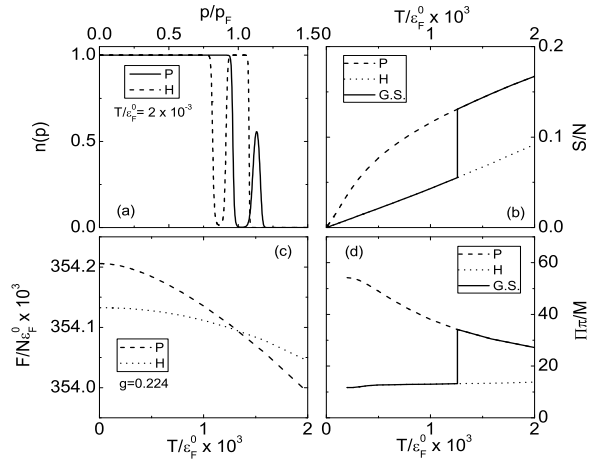}}
\caption{Panel (a): momentum distributions $n(p)$ for the ${\cal
P}$- and the ${\cal H}$-states at $T=2\cdot10^{-3}\varepsilon_F^0$.
Entropy per one particle $S/N$ (panel (b)), free energy $F$ per one
particle $F/N$ in units of $\varepsilon^0_F$ (panel (c)) and density
of single-particle states $\Pi$ in units of $M/\pi$ (panel (d)) as
functions of temperature $T$ in units of $\varepsilon^0_F$.
Parameters $g=0.224$, $q_0=2\,p_F$ are used in the  calculations. }
\label{fig:thermo}
\end{figure}

We return now to the model of 2D electron gas with the quasiparticle
interaction (\ref{model_2pf}). The height of the energy barrier
which, as seen in Fig.~\ref{fig:minima}, is of order
$10^{-3}\varepsilon^0_F$, determines the scale of temperature at
which on can expect a transition between ${\cal H}$- and ${\cal
P}$-states with increasing temperature. Such transition, indeed, occurs.
This is caused by the fact that the quasiparticle halo is
more narrow than the hole pocket, and hence, ``melts'' faster as
temperature increases, what is demonstrated in panel (a) of
Fig.~\ref{fig:thermo}. As a result, the entropy $S_{\cal P}(T)$ of
the ${\cal P}$-state increases faster with heating than the entropy
$S_{\cal H}(T)$ (see panel (b)), while the free energy $F_{\cal
P}(T)$ decreases more rapidly than $F_{\cal H}(T)$ (panel (c)).
Both minima equalize
at $T_1\simeq 1.2\cdot10^{-3}\varepsilon^0_F$, and the first order
transition occurs from the ${\cal H}$-state to the ${\cal P}$-state.
The entropy and the density of states undergo a jump at $T=T_1$ (see
panels (b) and (d)). The considered transition may have
relation to observed low-temperature anomalies in specific heat and
magnetic susceptibility of metals with heavy fermions.
\cite{Oeschler-PB-2008,Klingner-PRB-2011}

In conclusion, we analyzed the reconstruction of the Fermi surface
of the uniform Fermi system with increasing coupling constant of the
quasiparticle interaction and found that the topological transition,
in which two new connected sheets of the Fermi surface appear, is
followed by the
transition between two {\it topologically equivalent} states. The
Fermi surface of both these states consists of three connected
sheets, but one of these states, the ${\cal P}$-state, possesses a
structure of the quasiparticle halo, while the second one, the
${\cal H}$-state, that of the hole pocket. The transition from the
${\cal P}$-state to the ${\cal H}$-state is of the first order with
respect to the coupling constant $g$. As the temperature $T$
increases, the inverse first order transition from the ${\cal
H}$-state to the ${\cal P}$-state occurs due to more rapid
``melting'' of the narrow quasiparticle halo and more rapid increase
of its entropy with heating than increase of the entropy of the hole
pocket.

We thank V.~A.~Khodel and G.~E.~Volovik for fruitful discussions.
This research was supported by Grants No.~2.1.1/4540 and
NS-7235.2010.2 from the Russian Ministry of Education and Science,
and by Grant No.~09-02-01284 from the Russian Foundation for Basic
Research.

\end{document}